\documentstyle[10pt,twocolumn,psfig,amssymb,subeqn,cite]{IEEEtran}
\newcommand{\bi}{\begin{itemize}}
\newcommand{\ei}{\end{itemize}}

\newcommand{\be}{\begin{enumerate}}
\newcommand{\ee}{\end{enumerate}}
\newcommand{\bd}{\begin{description}}
\newcommand{\ed}{\end{description}}
\newcommand{\bc}{\begin{center}}
\newcommand{\ec}{\end{center}}
\newcommand{\bt}{\begin{tabbing}}
\newcommand{\et}{\end{tabbing}}
\newcommand{\bfig}{\begin{figure}}
\newcommand{\efig}{\end{figure}}
\newcommand{\beq}{\begin{equation}}
\newcommand{\beqarr}{\begin{eqnarray}}
\newcommand{\beqarrn}{\begin{eqnarray*}}
\newcommand{\eeq}{\end{equation}}
\newcommand{\eeqarr}{\end{eqnarray}}
\newcommand{\eeqarrn}{\end{eqnarray*}}
\newcommand{\bflr}{\begin{flushright}\vspace{-0.2in}}
\newcommand{\eflr}{\end{flushright}}
\newcommand{\bsub}{\begin{subequations}}
\newcommand{\esub}{\end{subequations}}
\newcommand{\barr}{\begin{array}}
\newcommand{\earr}{\end{array}}
\newcommand{\nn}{\nonumber}

\newcommand{\undb}[1]{\mbox{\bf{#1}}}

\begin{document}

\title{On the Sum of Correlated Squared $\kappa-\mu$ Shadowed Random Variables and its Application to
Performance Analysis of MRC \vspace*{-0.3em}}
%\author{
%\medskip 
%\normalsize 
%$\mbox{Manav R. Bhatnagar}$, {\em Member, IEEE} and Arti M.K., {\em Student Member, IEEE} \vspace*{-3em}
%\thanks{Manav R. Bhatnagar and Arti M.K. are
%with the Department of Electrical
%Engineering, Indian Institute of Technology - Delhi, Hauz Khas,
%New Delhi 110016, India (e-mail:
%{\tt \{manav,arti.mk\}@ee.iitd.ac.in}).}}
\author{
\medskip 
\normalsize 
$\mbox{Manav R. Bhatnagar}$, {\em Senior Member, IEEE}  \vspace*{-2.0em}
\thanks{Manav R. Bhatnagar is
with the Department of Electrical
Engineering, Indian Institute of Technology - Delhi, Hauz Khas,
New Delhi 110016, India (e-mail:
{\tt manav@ee.iitd.ac.in}).}
%\thanks{Copyright (c) 2013 IEEE. Personal use of this material is permitted. However, permission to use this material for any other purposes must be obtained from the IEEE by sending a request to pubs-permissions@ieee.org.}
}\date{}
%\markboth{S\lowercase{ubmitted to} \textit{IEEE Communications Letters} \lowercase{on} \today}
%{Bhardwaj and Mallik: 
%Capacity of Rake Reception with Energy Randomization
%} 

\setcounter{page}{1}
\maketitle

\begin{abstract}
In this paper, we study the statistical characterization %~\cite{king07}
of the sum of the squared $\kappa-\mu$ shadowed 
random
variables with correlated shadowing components. 
The probability density function (PDF) of this
sum is obtained in the form of a power series. The derived 
PDF is utilized for obtaining the performance results of
the maximal ratio combining (MRC) scheme over correlated
$\kappa-\mu$ shadowed 
fading channels. First, we derive 
the moment generating function (MGF) of the received
signal-to-noise ratio of the MRC receiver. By using the 
derived MGF expression,
the analytical diversity order is obtained; it is deduced
on the basis of this analysis that the diversity
of the MRC receiver over correlated $\kappa-\mu$ shadowed
channels
depends upon the number of diversity branches
and $\mu$ parameter. 
Further, the analytical average bit error rate of the
MRC scheme is also derived, which is applicable for 
$M$-PSK and $M$-QAM constellations. The Shannon
capacity of the correlated $\kappa-\mu$ shadowed 
channels is also derived in the 
form of the Meijer-G function. 
%Further, we also derive a power series based expression of the
%approximate capacity of the satellite system.    
\end{abstract}

\vspace*{-1em}
%%%%%%%%%%%%%%%%%%%%%%%%%%%%%%%%%%%%%%%%%%%%%%%%%%%%%%%%%%%%%%%%%%%%%%
\section{Introduction}
Fading is a well investigated propagation
phenomenon which has been researched extensively over the years.
The fading counts on several factors like the environment, scatterers,
rain, line-of-sight (LOS), snow,
the propagation frequency, etc. The fading severity
ranges from very
mild to extremely severe, depending upon the propagation
medium. 
A large number of models have been proposed in
the literature that reasonably well describe such a phenomenon
in its various aspects. 
Rayleigh, Hoyt, Weibull, Rice, Nakagami-$m$, and Shadowed-Rician
are the best known fading distributions very widely utilized
for the theoretical
studies of various practical wireless communication systems.
In~\cite{yacou07}, $\kappa-\mu$ fading is proposed
to allow flexibility to model the wireless channels fading fluctuations.
The $\kappa-\mu$ distribution is fully
characterized in terms of measurable physical parameters.
For many wireless communication systems, LOS is a dominating
factor which cannot be ignored. Moreover, in some wireless communication
systems, the LOS is not a deterministic parameter but its strength
randomly fluctuates over the time, e.g., in land mobile satellite (LMS)
links. For a general LOS propagation scenario, the $\kappa-\mu$ fading distribution 
provides a general multipath model. The $\kappa-\mu$ fading model is applicable to
some of the well studied classical fading models like one-sided Gaussian, Rayleigh,
Nakagami-$m$, and Rician fading. To be more precise, the fitting of the $\kappa-\mu$ 
distribution to the experimental data is better
than that attained by the conventional distributions mentioned before~\cite[Section V]{yacou07}.

The Shadowed-Rician channel model is used 
for modeling the LMS links because it 
yields significantly less computational burden as compared
to other LMS channel models~\cite{abdi03,bhatn13a,mk14a,bhatn13b,mk14b,bhatn14}.
This model shows nice fitting for the experimental channel measurements
under different shadowing conditions in LMS links~\cite{abdi03}. 
Here shadowing indicates that
the LOS component of the channel undergoes shadowing which
is modeled by a Nakagami-$m$ random variable (RV). The multipath fading
is characterized by the Rician fading in Shadowed-Rician fading model.
Since the $\kappa-\mu$ distribution includes the Rician distribution 
as a particular case, a natural
generalization of the $\kappa-\mu$ distribution can be obtained 
by a LOS shadow fading model with the 
same multipath/shadowing scheme used in the Shadowed-Rician model~\cite{paris14}. 
%The Shadowed-Rician fading is also a special case of the $\kappa-\mu$
%distribution. 
The statistical characterization of $\kappa-\mu$ shadowed 
fading is performed and analytical performance results for
the selection combining and maximal ratio combining (MRC) are derived under
independent LOS components in~\cite{paris14}. However, in the satellite communications
the LOS components associated with different spatial dimensions 
are not only dominating but are correlated with each other. The 
sum of \emph{correlated} squared Shadowed-Rician RVs and 
its application to communication systems performance
prediction is studied in~\cite{alfan07}.   
 
In this paper, we study the correlated $\kappa-\mu$ shadowed fading
for diversity reception technique. We statistically characterize the sum of correlated
squared $\kappa-\mu$ shadowed RVs in terms of probability density function (PDF)
 and moment generating
function (MGF). By using these characterizations, the error performance of the
MRC scheme is analyzed over the correlated $\kappa-\mu$ shadowed fading
channels. Specifically, we derive the average symbol error rate (SER)
and average bit error rate (BER) of the MRC receiver. We also derive the
analytical diversity order of the scheme and show that its diversity
is independent of the antenna correlations. Moreover, we find the ergodic capacity
of the correlated $\kappa-\mu$ shadowed channels, under MRC. 

 \section{Sum of Correlated $\kappa-\mu$ Shadowed Random Variables}  
The $\kappa-\mu$ distribution is a general model that describes various kinds of
fading encompassing from Rayleigh to Nakagami-$m$ fading;
it is useful for describing the small-scale variations along with
the LOS conditions. %The $\kappa-\mu$ shadowed fading studied in~\cite{paris14}
%is more general than the $\kappa-\mu$ fading proposed in~\cite{yacou07}. 
%The fading model of the $\kappa-\mu$ shadowed distribution~\cite{paris14}
The $\kappa-\mu$ fading
considers a signal composed of clusters of multipath waves, advancing in a 
non-homogenous environment. Within each cluster, the phases of the 
scattered multipath signals are random and have the same temporal delays.   
Moreover, a dominant component of arbitrary power is present in each cluster 
of the multipath waves having identical power. It is assumed that the intercluster delay-time
spreads are relatively larger than the delay times of the intracluster scattered
waves~\cite{yacou07}. 
The $\kappa-\mu$ shadowed fading proposed in~\cite{paris14}
is more general than the $\kappa-\mu$ fading proposed in~\cite{yacou07}. 
A general $\kappa-\mu$ shadowed fading model assumes that the 
dominant components of all clusters can randomly fluctuate due to
the shadowing. 

Let us consider a $\kappa-\mu$ shadowed distributed RV $X_l$, which is given by~\cite{paris14}
\beq 
X_l=\sum^{n_l}_{i=1}\left\{(W_{i,l}+jV_{i,l})+(\vartheta_l a_{i,l}+j\vartheta_l b_{i,l})\right\},
\label{k-m distr}
\eeq 
where $W_{i,l}$ and $V_{i,l}$ are mutually independent zero mean Gaussian RVs with $\sigma^2$ 
variance; $n_l$ is a natural number, $a_{i,l}$ and $b_{i,l}$ are real numbers;
and $\vartheta_l$ is a Nakagami-$m$ distributed RV with shaping parameter $m$ and
spreading parameter $\Omega_l=1$. In (\ref{k-m distr}), $W_{i,l}+jV_{i,l}$, which
is circularly symmetric complex Gaussian RV, represents
the scattered component of the $i$-th cluster. %corresponding
%to the small scale fading component of $2\sigma^2$ power. 
On the other hand $\vartheta_l a_{i,l}+j\vartheta_l b_{i,l}$ denotes the
dominating LOS component with $a^2_{i,l}+b^2_{i,l}$ power. 
The common shadowing fluctuation of all clusters is represented by the power-normalized
RV $\vartheta_l$; for deterministic LOS case, $\vartheta_l=1$.  

From~(\ref{k-m distr}), it can be easily shown that the power of $X_l$, i.e., $Y_l=X^2_l$ is 
given by~\cite{paris14}
\beq
Y_l=\sum^{n_l}_{i=1}\left\{(W_{i,l}+\vartheta_l a_{i,l})^2+(V_{i,l}+\vartheta_l b_{i,l})^2\right\}.
\label{powrk-m} 
\eeq
The average value of $Y_l$ is $E[Y_l]=2n_l\sigma^2+\sum^{n_l}_{i=1}(a^2_{i,l}+b^2_{i,l})$ as seen
from (\ref{powrk-m}); here the expectation is denoted by $E[\cdot]$. 
It can be seen from (\ref{powrk-m}) that conditioned on $\vartheta_l$, $Y_l$ is sum of
$2n_l$ independent non-central Chi-squared distributed RVs; therefore, the conditional PDF
of $Y_l$ will be~\cite{proak01}
\beqarr
f_{Y_l|\vartheta_l}(y)&=&\frac{1}{2\sigma^2}\left(\frac{y}{\vartheta^2_l\sum^{n_l}_{i=1}(a^2_{i,l}+b^2_{i,l})}\right)^{\frac{n_l-1}{2}}\nn\\
&\times&e^{-\frac{y+\vartheta^2_l\sum^{n_l}_{i=1}(a^2_{i,l}+b^2_{i,l})}{2\sigma^2}}\nn\\
&\times&I_{n_l-1}\left(\frac{\vartheta_l\sqrt{\sum^{n_l}_{i=1}(a^2_{i,l}+b^2_{i,l})}}{\sigma^2}\sqrt{y}\right),
\label{condk-m}
\eeqarr
where $I_\nu(\cdot)$ is the modified Bessel function of the first kind. 
Let us now define the following substitution variables: 
\beqarr
\sum^{n_l}_{i=1}(a^2_{i,l}+b^2_{i,l})&\triangleq& 2\sigma^2\kappa_l\mu_l\nn\\
n_l&\triangleq&\mu_l.
\label{sub}
\eeqarr
Further, let us also define a new RV $\gamma_l\triangleq\bar{\gamma}Y_l/E[Y_l]$, denoting the instantaneous signal-to-noise ratio (SNR) 
for the signal received under the $\kappa-\mu$ shadowed fading with $\bar{\gamma}$ being the average SNR. By using
the method of transformation of RVs~\cite{proak01} along with substituting the new variables given in (\ref{sub}), in (\ref{condk-m}), we get the conditional PDF of $\gamma_l$:
\beqarr
f_{\gamma_l|\vartheta_l}(\gamma)&=&\frac{\mu_l(1+\kappa_l)^{\frac{\mu_l+1}{2}}}{\bar{\gamma}^{\frac{\mu_l+1}{2}}\kappa^{\frac{\mu_l-1}{2}}_l{\vartheta^{\mu_l-1}_l}}
{\gamma}^{\frac{\mu_l-1}{2}}\nn\\
&\times& e^{-\frac{\mu_l(1+\kappa_l)\gamma}{\bar{\gamma}}-\vartheta^2_l\mu_l\kappa_l}\nn\\
&\times& I_{\mu_l-1}\left(2\mu_l\vartheta_l\sqrt{\frac{\kappa_l(1+\kappa_l)\gamma}{\bar{\gamma}}}\right).
\label{condk-m1}
\eeqarr
Let us now define the sum of $L$ squared $\kappa-\mu$ shadowed RVs as 
\beq
Y\triangleq\sum^L_{l=1}Y_l. 
\label{sum}
\eeq
 We encounter this sum in the diversity reception schemes like MRC. From (\ref{powrk-m}) and (\ref{sum}), it can be inferred that $Y$ is a sum of $2\sum^L_{l=1}n_l$ non-central Chi-squared RVs. Hence, the conditional PDF of $Y$ will be given by~\cite{proak01}
\beqarr
&&\hspace*{-1.5em}f_{Y|\vartheta_1,..,\vartheta_l}(y)=\frac{1}{2\sigma^2}\!\!\left(\frac{y}{\sum^L_{l=1}\vartheta^2_l[\sum^{n_l}_{i=1}(a^2_{i,l}+b^2_{i,l})]}\right)^{\frac{\sum^L_{l=1}n_l-1}{2}}\nn\\
&&\hspace*{0em}\times e^{-\frac{\sum^L_{l=1}\vartheta^2_l[\sum^{n_l}_{i=1}(a^2_{i,l}+b^2_{i,l})]+y}{2\sigma^2}}\nn\\
&&\hspace*{0em}\times I_{\sum^L_{l=1}n_l-1}\left(\frac{\sqrt{\sum^L_{l=1}\vartheta^2_l[\sum^{n_l}_{i=1}(a^2_{i,l}+b^2_{i,l})]}}{\sigma^2}\sqrt{y}\right).
\eeqarr
For the diversity reception scheme, the instantaneous SNR is defined as $\gamma\triangleq y\bar{\gamma}/E[Y]$~\cite{paris14}, where the average value of $Y$, i.e., $E[Y]$ is given by $\sum^L_{l=1}\sum^{n_l}_{i=1}(a^2_{i,l}+b^2_{i,l})+2\sigma^2\sum^L_{l=1}n_l$. By employing the substitution of variables given in (\ref{sub}) and using the method of transformation of RVs, we get the conditional PDF of $\gamma$ as
\beqarr
&&\hspace*{-2.5em}f_{\gamma|\vartheta^2_1,..,\vartheta^2_L}(\gamma)=\frac{(\sum^L_{l=1}\mu_l(1+\kappa_l))^{\frac{\sum^L_{l=1}\mu_l+1}{2}}}{\bar{\gamma}^{\frac{\sum^L_{l=1}\mu_l+1}{2}}(\sum^L_{l=1}\vartheta^2_l\mu_l\kappa_l)^{\frac{\sum^L_{l=1}\mu_l-1}{2}}}\nn\\
&&\hspace*{-2.5em}\times
{\gamma}^{\frac{\sum^L_{l=1}\mu_l-1}{2}}e^{-\frac{\sum^L_{l=1}\mu_l(1+\kappa_l)\gamma}{\bar{\gamma}}-\sum^L_{l=1}\vartheta^2_l\mu_l\kappa_l}\nn\\
&&\hspace*{-2.5em}\times I_{\sum^L_{l=1}\mu_l-1}\!\!\left(\!\!2\sqrt{\!\!\Bigg(\sum^L_{l=1}\vartheta^2_l\mu_l\kappa_l\!\!\Bigg)\!\Bigg(\!\!\sum^L_{l=1}\mu_l(1+\kappa_l)\!\!\Bigg)}\sqrt{\frac{\gamma}{\bar{\gamma}}}\right)\!\!.
\label{condk-m2}
\eeqarr
In (\ref{condk-m2}), let us use another substitution $\vartheta^2_l\mu_l\kappa_l=\tilde{\vartheta}^2_l$ and have
\beqarr
&&\hspace*{-2.5em}f_{\gamma|\vartheta^2_1,..,\vartheta^2_L}(\gamma)=\frac{(\sum^L_{l=1}\mu_l(1+\kappa_l))^{\frac{\sum^L_{l=1}\mu_l+1}{2}}}{\bar{\gamma}^{\frac{\sum^L_{l=1}\mu_l+1}{2}}(\sum^L_{l=1}\tilde{\vartheta}^2_l)^{\frac{\sum^L_{l=1}\mu_l-1}{2}}}\nn\\
&&\hspace*{-2.5em}\times
{\gamma}^{\frac{\sum^L_{l=1}\mu_l-1}{2}}e^{-\frac{\sum^L_{l=1}\mu_l(1+\kappa_l)\gamma}{\bar{\gamma}}-\sum^L_{l=1}\tilde{\vartheta}^2_l}\nn\\
&&\hspace*{-2.5em}\times I_{\sum^L_{l=1}\mu_l-1}\!\!\left(\!\!2\sqrt{\Bigg(\sum^L_{l=1}\tilde{\vartheta}^2_l\Bigg)\Bigg(\sum^L_{l=1}\mu_l(1+\kappa_l)\Bigg)}\sqrt{\frac{\gamma}{\bar{\gamma}}}\right).
\label{condk-m3}
\eeqarr
In (\ref{condk-m3}), $\sum^L_{l=1}\tilde{\vartheta}^2_l$ denotes the sum of $L$ Gamma distributed RVs; the shape parameter of $\tilde{\vartheta}^2_l$ is $m$ and scale parameter is $\mu_l\kappa_l/m$. 

Let us assume that the dominating components of 
the $\kappa-\mu$ shadowed RVs, i.e., $\tilde{\vartheta}^2_l$ are correlated with $\rho_{ij}$ correlation coefficient given by
\beq
\rho_{ij}=\rho_{ji}=\frac{\mbox{Cov}(\tilde{\vartheta}^2_i,\tilde{\vartheta}^2_j)}{\sqrt{\mbox{Var}(\tilde{\vartheta}^2_i)\mbox{Var}(\tilde{\vartheta}^2_j)}}, \:\: 0\leq \rho_{ij}\leq 1,\:\: i,j=1,..,L.
\label{corrcoef}
\eeq
Then the PDF of $Z=\sum^L_{l=1}\tilde{\vartheta}^2_l$ can be expressed as~\cite{aloun01}
\beq
f_Z(z)=\prod^L_{l=1}\left(\frac{\lambda_1}{\lambda_l}\right)^m\sum^\infty_{k=0}\frac{\delta_kz^{Lm+k-1}e^{-z/\lambda_1}}{\lambda^{Lm+k}_1\Gamma(Lm+k)} u(z),
\label{corr}
\eeq
where $\lambda_1=\mbox{min}_l\left\{\lambda_l\right\}$, $\left\{\lambda_l\right\}^L_{l=1}$ are the eigenvalues of the matrix $\undb{DC}$,
$\undb{D}$ being a diagonal matrix with the entries $\left\{\mu_l\kappa_l/m\right\}^L_{l=1}$ and $\undb{C}$ is the $L\times L$ positive definite matrix defined by
\beqarr
\undb{C}\triangleq\left[\begin{array}{cccc}1&\sqrt{\rho_{12}}&\cdots& \sqrt{\rho_{1L}}\\
\sqrt{\rho_{21}}&1&\cdots& \sqrt{\rho_{2L}}\\
\cdot&\cdot&\cdot&\cdot\\
\cdot&\cdot&\cdot&\cdot\\
\sqrt{\rho_{L1}}&\cdots&\cdots&1
\end{array}\right]_{L\times L},
\eeqarr
and the coefficients $\delta_k$ can be obtained recursively as
\beqarr
\delta_{k+1}=\frac{m}{k+1}\sum^{k+1}_{i=1}\left[\sum^L_{j=1}\left(1-\frac{\lambda_1}{\lambda_j}\right)^i\right]\delta_{k+1-i},\;\; k=0,1,..
\label{del}
\eeqarr
with $\delta_0=1$. 
Note that the PDF of (\ref{corr}) is in the form of a power series. It is shown in~\cite{aloun01,mosch85} that
it is a converging power series. Further, a similar correlation
model is also considered for the Shadowed-Rician LMS channels in~\cite{alfan07,dhung12}, where
the shadowing components are correlated and multipath components are uncorrelated.  

\begin{figure}[t!]\vspace*{-1em}
\hspace*{-0.0em}
\centerline{\psfig{file=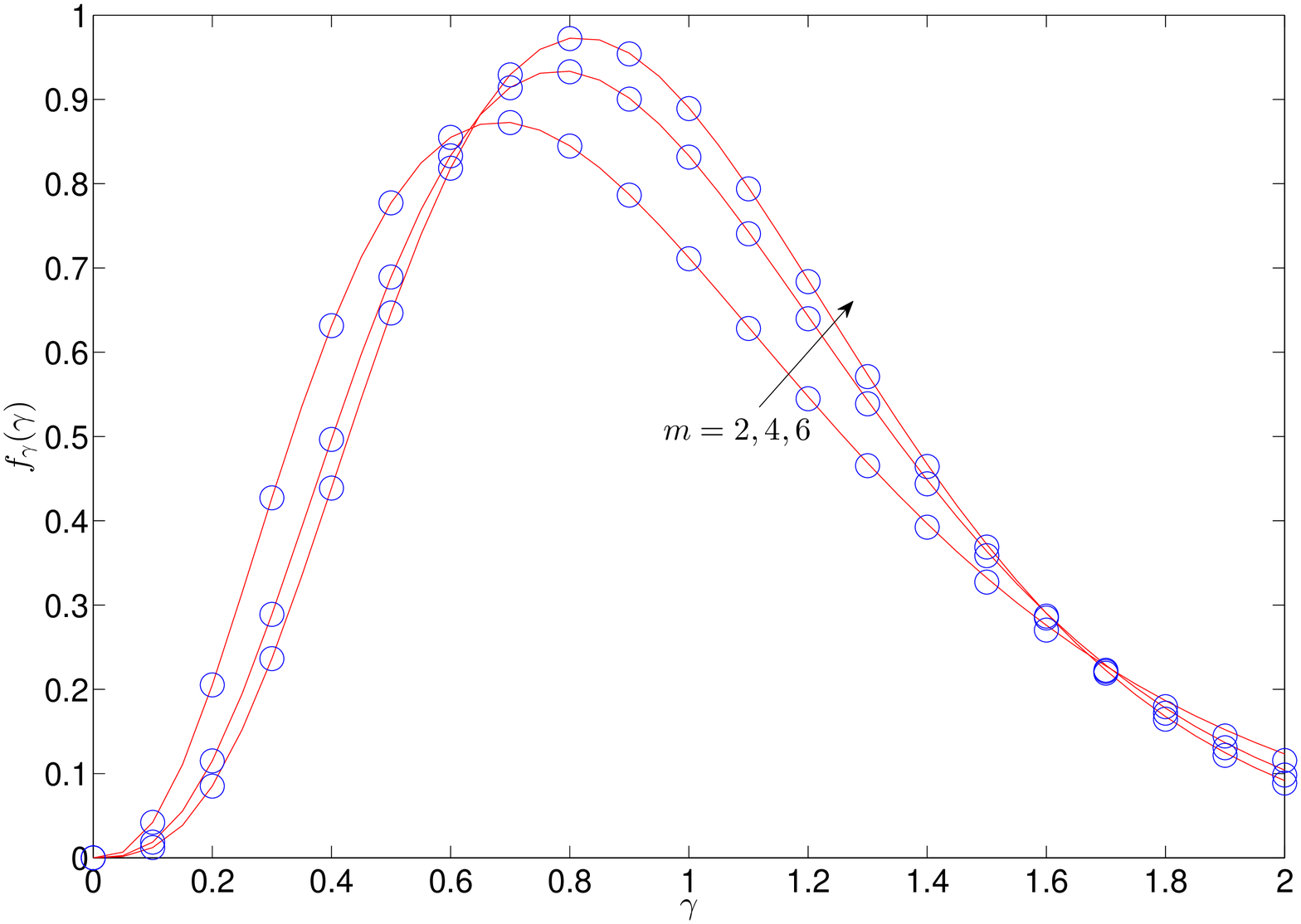,width=3.8in,height=2.5in}}\vspace*{-4mm}
\caption{Analytical $-\!\!-$ and simulated $\circ$ PDFs for $\kappa_l=2$, $\mu_l=2$, $m=2,4,6$, $L=2$, and $\rho_{ij}=0.7^{|i-j|}$.}\vspace*{-6mm}
\label{fig1}
\end{figure}

If $\epsilon=\sum^L_{l=1}\mu_l$ and $\eta=\sum^L_{l=1}\mu_l(1+\kappa_l)$, then the conditional PDF of $\gamma$ can be very compactly represented by
\beqarr
f_{\gamma|z}(\gamma)=\frac{\eta^{\frac{\epsilon+1}{2}}\gamma^{\frac{\epsilon-1}{2}}}{\bar{\gamma}^{\frac{\epsilon+1}{2}}e^{\frac{\eta}{\bar{\gamma}}\gamma}}z^{-\frac{\epsilon-1}{2}}e^{-z}I_{\epsilon-1}\left(\sqrt{\frac{4\gamma\eta}{\bar{\gamma}}}\sqrt{z}\right).
\label{comp}
\eeqarr
For finding the unconditional PDF of $\gamma$, (\ref{comp}) should be averaged upon $Z$; therefore, from (\ref{corr}) and (\ref{comp}), we get
\beqarr
&&\hspace*{-2.5em}f_{\gamma}(\gamma)=A\left(\frac{\eta}{\bar{\gamma}}\right)^{\frac{\epsilon+1}{2}}\gamma^{\frac{\epsilon-1}{2}} e^{-\frac{\eta}{\bar{\gamma}}\gamma}\sum^\infty_{k=0}D_k\nn\\
&&\hspace*{-2.5em}\times\int^\infty_{0}z^{Lm+k-\frac{\epsilon}{2}-\frac{1}{2}}e^{-z\left(1+\frac{1}{\lambda_1}\right)}I_{\epsilon-1}\left(\sqrt{\frac{4\gamma\eta}{\bar{\gamma}}}\sqrt{z}\right)dz,
\label{int}
\eeqarr
where 
$A=\prod^L_{l=1}\left(\frac{\lambda_1}{\lambda_l}\right)^m$ and $D_k=\frac{\delta_k}{\lambda^{Lm+k}_1\Gamma(Lm+k)}$.
The integral in (\ref{int}) can be solved by using~\cite[Eq. (2.15.5.4)]{prudn92}, and the unconditional PDF of $\gamma$
can be written as
\beqarr
&&\hspace*{-3em}f_{\gamma}(\gamma)=A\left(\frac{\eta}{\bar{\gamma}}\right)^{\epsilon}\gamma^{{\epsilon-1}} e^{-\frac{\eta}{\bar{\gamma}}\gamma}\sum^\infty_{k=0}\tilde{D}_k\nn\\
&&\times {}_1F_1\big(Lm+k;\epsilon;\frac{\eta\gamma}{\bar{\gamma}\left(1+\frac{1}{\lambda_1}\right)}\Big),
\label{final}
\eeqarr
where
\beq
\tilde{D}_k=\frac{\delta_k}{\lambda^{Lm+k}_1\Gamma(\epsilon)}\left(1+\frac{1}{\lambda_1}\right)^{-(Lm+k)}
\eeq
 and ${}_1F_1(a;b;z)$ is the confluent Hypergeometric function~\cite{grand00}.
The derived expression of the PDF (\ref{final}) is utilized for plotting the analytical values of the PDF
in Fig.~\ref{fig1} for $\kappa_l=2$, $\mu_l=2$, $m=2,4,6$, $L=2$, and exponential correlation $\rho_{ij}=0.7^{|i-j|}$. In addition,
the simulated PDF is also shown for these parameter values in the figure. It can be seen
from the figure that the simulated and analytical PDFs are closely matched. This justifies
the correctness of the derived PDF. Further, it can be seen from the figure that the PDF plot
becomes peaky with increasing value of $m$. 
\subsection{Sum of I.I.D. Squared $\kappa-\mu$ Shadowed Random Variables}
If all squared $\kappa-\mu$ shadowed RVs are independent and identically distributed (i.i.d.), then $\kappa_l=\kappa$
and $\mu_l=\mu$. From (\ref{corrcoef})-(\ref{del}), it can be found
that in this case, $\rho_{ij}=\rho_{ji}=0$ and the matrix $\undb{DC}$ contains uniform eigenvalues,
$\lambda_l=\kappa\mu/m$; hence, $\delta_k=0$, $\forall k>0$. After substituting these values in
(\ref{final}), we get the PDF of sum of i.i.d. squared $\kappa-\mu$ shadowed RVs:
\beqarr
&&\hspace*{-3em}f_{\gamma}(\gamma)=\left(\frac{L\mu(1+\kappa)}{\bar{\gamma}}\right)^{L\mu}\left(\frac{m}{m+\kappa\mu}\right)^{Lm}\frac{\gamma^{L\mu-1}}{\Gamma(L\mu)}\nn\\
&&\hspace*{-3em}\times e^{-\frac{L\mu(1+\kappa)}{\bar{\gamma}}\gamma}{}_1F_1\left(Lm;L\mu;\frac{L\kappa\mu^2(1+\kappa)\gamma}{\bar{\gamma}(m+\kappa\mu)}\right).
\label{finaliid}
\eeqarr
It can be easily verified from (\ref{finaliid}) that for $L=1$, we get the PDF of the square of
a single $\kappa-\mu$ shadowed RV which matches with the PDF given in \cite[Eq. (4)]{paris14}. 
\section{Performance Analysis of the MRC Diversity System}
\subsection{System and Channel Model}
Consider an $L$-branch MRC at the receiver, where each branch 
experiences the correlated $\kappa-\mu$ shadowed fading with an instantaneous
SNR $\gamma_l$, $l=1,..,L$; the correlation coefficient is
given in (\ref{corrcoef}). The received instantaneous SNR of the
MRC receiver, i.e., $\gamma=\sum^L_{l=1}\gamma_l$ is characterized in (\ref{final}).

\subsection{Moment Generating Function of the Received SNR}
The MGF of the received SNR is expressed as
\beq
M_\gamma(s)=E_{\gamma}[e^{-s\gamma}].
\label{mgfbasic}
\eeq
%where $E[\cdot]$ denotes the expectation. 
From~(\ref{final}) and (\ref{mgfbasic}), the MGF of the SNR will be
\beqarr
&&\hspace*{-3em}M_\gamma(s)=A\left(\frac{\eta}{\bar{\gamma}}\right)^{\epsilon}\sum^\infty_{k=0}\tilde{D}_k\nn\\
&&\hspace*{-3em}\times
\int^\infty_0\gamma^{{\epsilon-1}} e^{-\left(s+\frac{\eta}{\bar{\gamma}}\right)\gamma}{}_1F_1\big(Lm+k;\epsilon;\frac{\eta\gamma}{\bar{\gamma}\left(1+\frac{1}{\lambda_1}\right)}\Big)d\gamma.
\label{mgfint}
\eeqarr
The following relations can be employed in (\ref{mgfint}):
\beqarr
&&\hspace*{-2em}{}_1F_1\Big(Lm+k;\epsilon;\frac{\eta\gamma}{\bar{\gamma}\left(1+\frac{1}{\lambda_1}\right)}\Big)\nn\\
&&\hspace*{-2em}
=\frac{\Gamma\left(\epsilon\right)}{\Gamma\left(Lm+k\right)}G^{11}_{12}\Bigg(-\frac{\eta\gamma}{\bar{\gamma}\left(1+\frac{1}{\lambda_1}\right)}\Bigg|{\hspace*{-0em}1-Lm-k\atop{0,1-\epsilon}}\Bigg)
\label{hypgrel}
\eeqarr
 and
   \beqarr
e^{-(s+\frac{\eta}{\bar{\gamma}})\gamma}=G^{10}_{01}\Bigg((s+\frac{\eta}{\bar{\gamma}})\gamma\Bigg|{.\atop{0}}\Bigg),
\label{exp}
\eeqarr 
where $G_{p,q}^{m,n}(\cdot|{\cdots\atop \cdots})$ is the Meijer-G function~\cite[Eq.~(9.301)]{grand00}. After substitution of these relations, we get
\beqarr
&&\hspace*{-3em}M_\gamma(s)=A\left(\frac{\eta}{\bar{\gamma}}\right)^{\epsilon}\sum^\infty_{k=0}\tilde{D}_k\frac{\Gamma\left(\epsilon\right)}{\Gamma\left(Lm+k\right)} 
\int^\infty_0\gamma^{{\epsilon-1}}\nn\\
&&\hspace*{-3em} \times G^{10}_{01}\Bigg(\!\!(s+\frac{\eta}{\bar{\gamma}})\gamma\Bigg|{.\atop{0}}\!\!\Bigg)G^{11}_{12}\Bigg(\!\!\frac{-\eta\gamma}{\bar{\gamma}\left(1+\frac{1}{\lambda_1}\right)}\Bigg|{\hspace*{-0em}1-Lm-k\atop{0,1-\epsilon}}\!\!\Bigg)d\gamma.
\label{mgfint1}
\eeqarr
\begin{figure}[t!]\vspace*{-1em}
\hspace*{-0.5em}
\centerline{\psfig{file=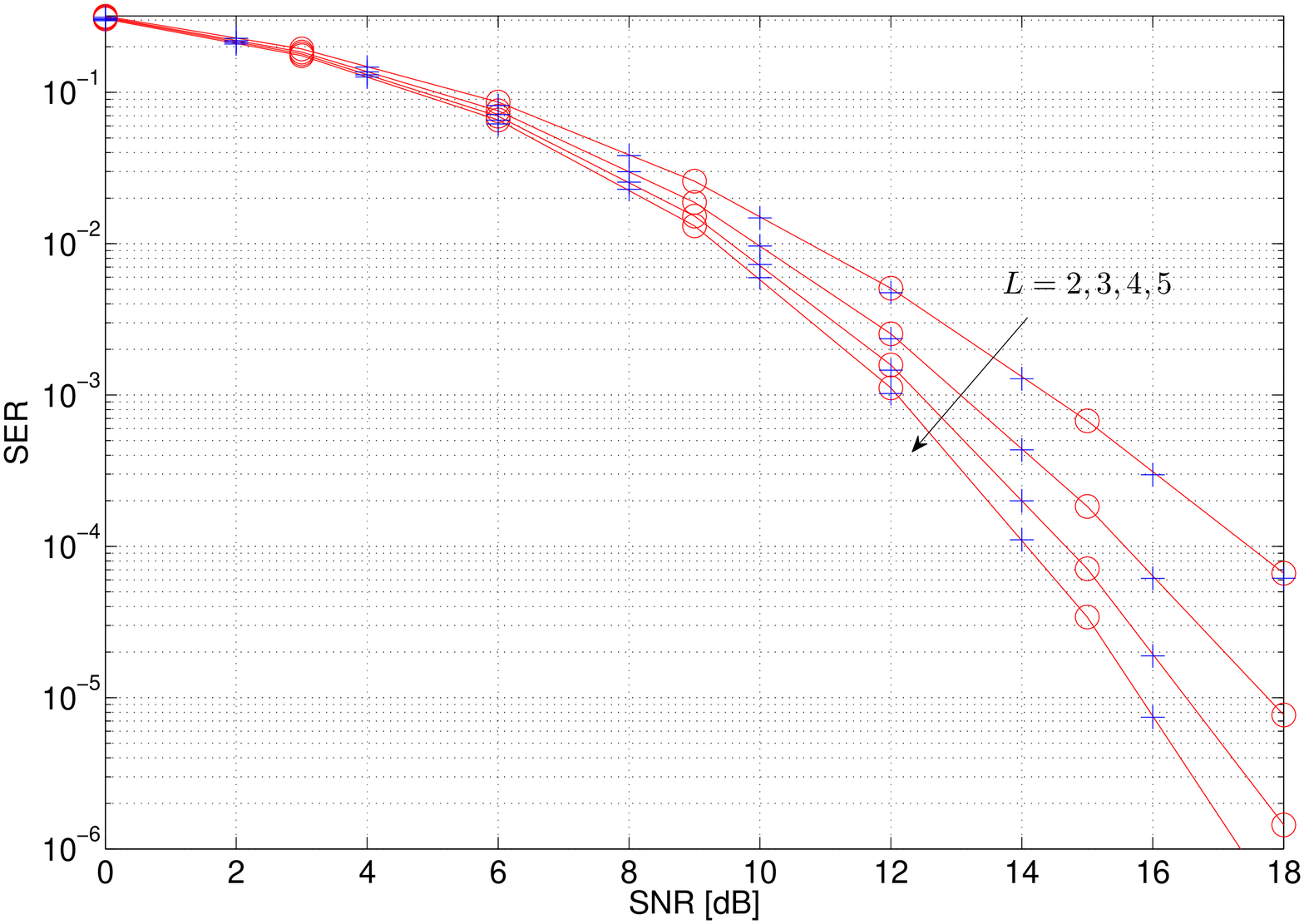,width=4in,height=3.0in}}\vspace*{-4mm}
\caption{Analytical $-\circ-$ and simulated $+$ SERs for $\kappa_l=5$, $\mu_l=2$, $m=2$, $\rho_{ij}=0.5^{|i-j|}$, $L=2,3,4,5$, and QPSK constellation.}\vspace*{-6mm}
\label{fig2}
\end{figure}
The integral in (\ref{mgfint1}) can be solved by using \cite[Eq. (21)]{adamc90} and we get
\beqarr
&&\hspace*{-3em}M_\gamma(s)=A\left(\frac{\eta}{\bar{\gamma}}\right)^{\epsilon}\sum^\infty_{k=0}\tilde{D}_k\frac{\Gamma\left(\epsilon\right)}{\Gamma\left(Lm+k\right)}\left(s+\frac{\eta}{\bar{\gamma}}\right)^{-\epsilon}\nn\\
&&\hspace*{-3em}\times G^{12}_{22}\Bigg(\frac{-\eta}{\bar{\gamma}\left(1+\frac{1}{\lambda_1}\right)\left(s+\frac{\eta}{\bar{\gamma}}\right)}\Bigg|{\hspace*{-0em}1-Lm-k,1-\epsilon\atop{0,1-\epsilon}}\Bigg).   
\label{mgfint2}
\eeqarr
The MGF for i.i.d. case can be easily calculated by using (\ref{finaliid}) and method given above. Alternatively,
the MGF for i.i.d. case can also be obtained from (\ref{mgfint2}) by putting $\delta_0=1$, $\delta_k=0$, $k>0$, $\lambda_l=\kappa\mu/m$, $\epsilon=L\mu$, and $\eta=L\mu(1+\kappa)$. 

The SER of the scheme for $M$-PSK constellation can be efficiently calculated by employing the following relation~\cite[Eq. (10)]{mckay09}:
\beqarr
P_{MPSK}\approx\sum^3_{p=1}\beta_pM_{\gamma}\left(\alpha_p\right),
\label{intser1}
\eeqarr
where $\beta_1 = \theta_M/(2\pi) - 1/6$, $\beta_2 = 1/4$, $\beta_3 = \theta_M/(2\pi) - 1/4$, $\alpha_1=g_{MPSK}$, $\alpha_2=4g_{MPSK}/3$, $\alpha_3=g_{MPSK}/sin^2(\theta_M)$, $g_{MPSK}=sin^2(\pi/M)$, and $\theta_M=(M-1)\pi/M$.

The analytical SER for $\kappa_l=5$, $\mu_l=2$, $m=2$, $\rho_{ij}=0.5^{|i-j|}$, $L=2,3,4,5$, and QPSK constellation is plotted in Fig.~\ref{fig2}. The simulated SER versus SNR plots are also shown for the same parameters in the figure. The simulated SER values are obtained by using $10^7$ channel realizations. The simulated SER closely follows the analytical SER values at all SNR values considered in the figure. Further, the performance of the MRC receiver improves with increasing value of antennas, as seen from the figure.  

\begin{figure}[t!]\vspace*{-1em}
\hspace*{-0.0em}
\centerline{\psfig{file=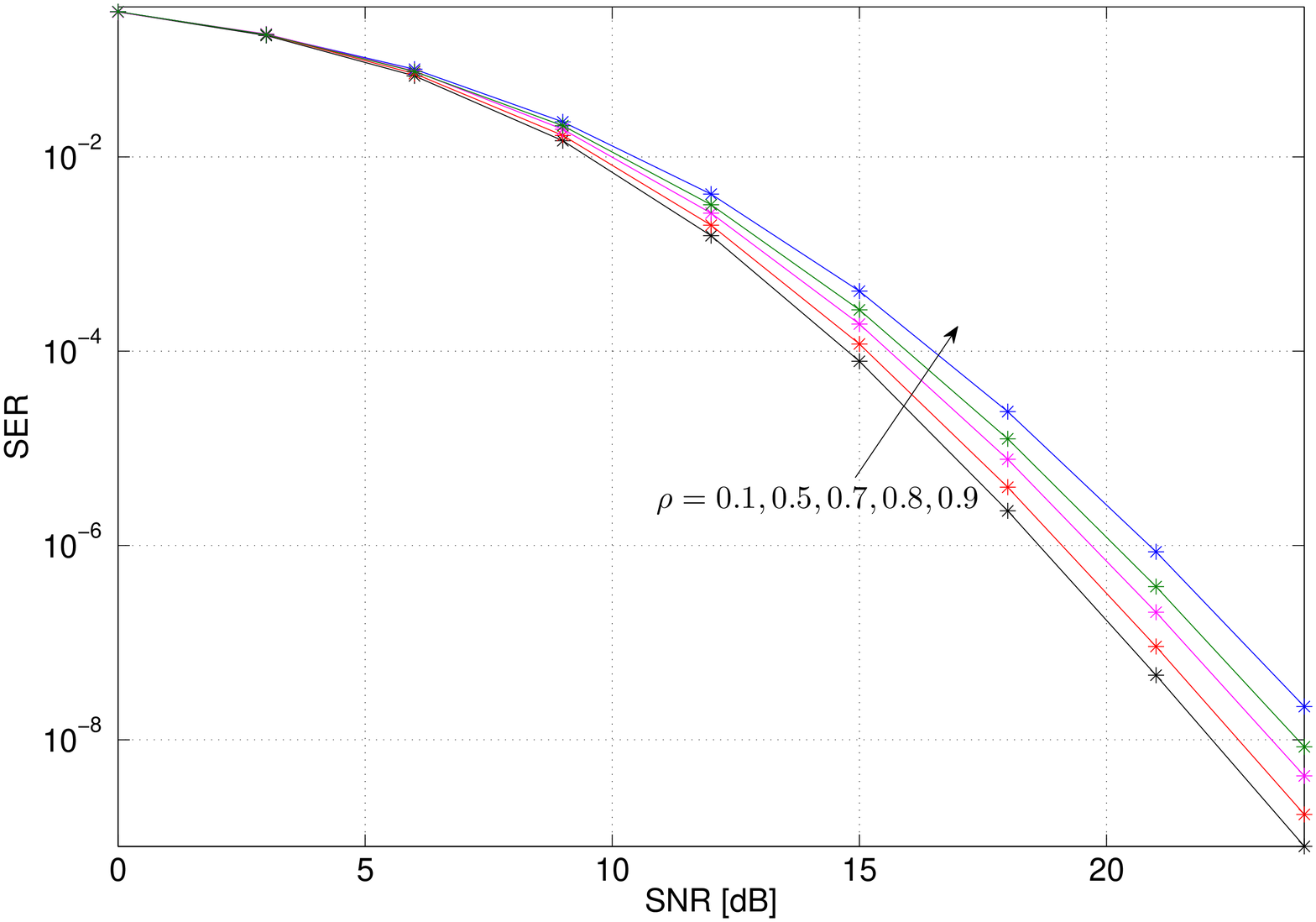,width=4in,height=3.0in}}\vspace*{-4mm}
\caption{Analytical SER versus SNR plots for $\kappa_l=5$, $\mu_l=2$, $m=2.5$, $\rho_{ij}=\rho^{|i-j|}$, $\rho=0.1,0.5,0.7,0.8,0.9$, $L=3$, and QPSK constellation.}\vspace*{-6mm}
\label{fig5}
\end{figure}
The effect of correlation parameter $\rho$ on the SER performance of the MRC receiver is shown in Fig.~\ref{fig5}
for $\kappa_l=5$, $\mu_l=2$, $m=2.5$, $\rho_{ij}=\rho^{|i-j|}$, $\rho=0.1,0.5,0.7,0.8,0.9$, $L=3$, and QPSK constellation. 
It can be seen from the figure that the receiver error performance degrades with increasing value of the correlation coefficient.
\subsection{Diversity Order Calculation}
By using the Slater's theorem~\cite[Eq.~(8.2.2.3)]{prudn90} which represents the Meijer-G function as a finite series of the Hypergeometric function, it can be shown that
\beqarr
&&\hspace*{-1em}G^{12}_{22}\Bigg(\frac{-\eta}{\bar{\gamma}\left(1+\frac{1}{\lambda_1}\right)\left(s+\frac{\eta}{\bar{\gamma}}\right)}\Bigg|{\hspace*{-0em}1-Lm-k,1-\epsilon\atop{0,1-\epsilon}}\Bigg)\nn\\
&&\hspace*{-1em}=\Gamma(Lm+k){}_2F_1\!\!\!\left(Lm+k,\epsilon;\epsilon;\frac{\eta\left(s+\frac{\eta}{\bar{\gamma}}\right)^{-1}}{\bar{\gamma}\left(1+\frac{1}{\lambda_1}\right)}\right),
\label{stal}
\eeqarr 
where ${_2F_1(a_1,a_2;b_1;z)}$ is the Gaussian Hypergeometric function~\cite{grand00}. From~(\ref{mgfint2}) and (\ref{stal}),
the MGF of the MRC scheme can be written as
\beqarr
&&\hspace*{-3em}M_\gamma(s)=A\left(\frac{\eta}{\bar{\gamma}}\right)^{\epsilon}\sum^\infty_{k=0}\tilde{D}_k{\Gamma\left(\epsilon\right)}\left(s+\frac{\eta}{\bar{\gamma}}\right)^{-\epsilon}\nn\\
&&\hspace*{-3em}\times {}_2F_1\left(Lm+k,\epsilon;\epsilon;\frac{\eta}{\bar{\gamma}\left(1+\frac{1}{\lambda_1}\right)\left(s+\frac{\eta}{\bar{\gamma}}\right)}\right).   
\label{mgfint3}
\eeqarr
For diversity calculation, let us assume that $\bar{\gamma}$ is very large, which means that ${\eta}/\left({\bar{\gamma}\left(1+\frac{1}{\lambda_1}\right)\left(s+\frac{\eta}{\bar{\gamma}}\right)}\right)$ is very small. 
Therefore, after observing the fact that ${_2F_1(a_1,a_2;b_1;z)}\rightarrow 1$, $z\rightarrow 0$~\cite{abram72} and after some other algebra, the asymptotic
value of the MGF is given by
\beqarr
&&\hspace*{-3em}M_\gamma(s)=A\left(\frac{\eta}{\bar{\gamma}}\right)^{\epsilon}\sum^\infty_{k=0}\tilde{D}_k{\Gamma\left(\epsilon\right)}s^{-\epsilon}.   
\label{mgfasymp1}
\eeqarr
The diversity order of the MRC scheme is $\epsilon=\sum^L_{l=1}\mu_l$, as seen from (\ref{mgfasymp1}).  

\begin{figure}[t!]\vspace*{-1em}
\hspace*{0.0em}
\centerline{\psfig{file=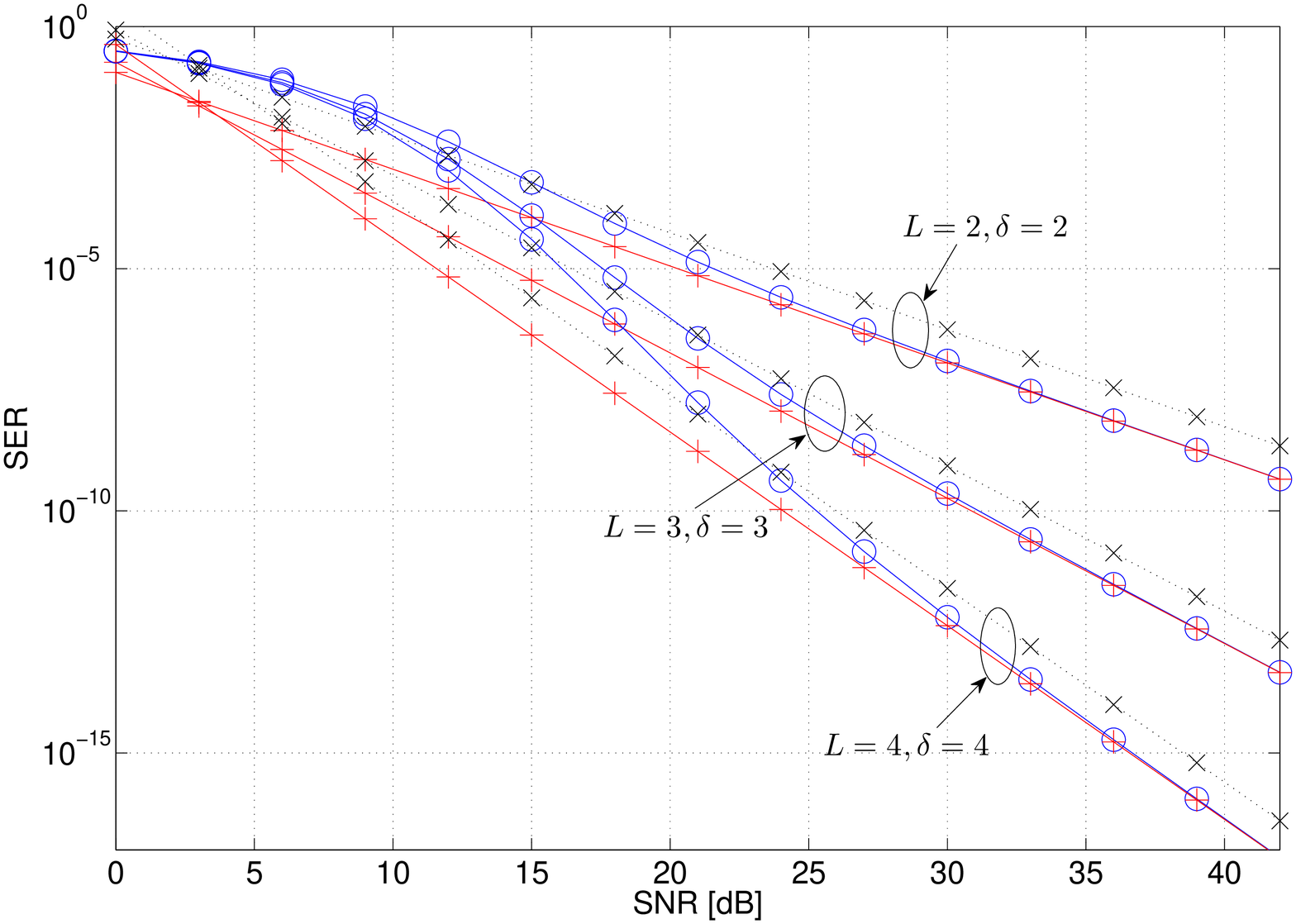,width=4in,height=3.0in}}\vspace*{-4mm}
\caption{Analytical $-\circ-$ and asymptotic $-+-$ SERs for $\kappa_l=10$, $\mu_l=1$, $m=3$, $\rho_{ij}=0.1^{|i-j|}$, $L=2,3,4$, and QPSK constellation; the asymptotic ideal diversity plots are shown by $-\times-$. }\vspace*{-6mm}
\label{fig3}
\end{figure}
For verifying the analytical diversity order, we have plotted the asymptotic SER by using (\ref{intser1}) and (\ref{mgfasymp1}) in Fig.~\ref{fig3}. The analytical SER versus SNR plots (obtained from (\ref{mgfint2}) and (\ref{intser1})) are also shown in the figure. 
All plots are given for $\kappa_l=10$, $\mu_l=1$, $m=3$, $\rho_{ij}=0.1^{|i-j|}$, $L=2,3,4$, and QPSK constellation. It can be seen from the figure that the asymptotic SER closely overlaps with the analytical SER at high values of the SNR. Therefore, the proposed
asymptotic MGF in (\ref{mgfasymp1}) is sufficiently tight at high SNR. Further, we have also plotted the asymptotic ideal diversity curves by using the relation: $\alpha/\bar{\gamma}^{\delta}$, where $\alpha$ is a positive real coefficient and $\delta$ denotes the ideal diversity. The SER versus SNR performance of the MRC receiver decays at the same rate as that of the slope of the asymptotic ideal diversity curves in all cases. 
The correctness of the diversity analysis is corroborated from the figure. 
 
\subsection{Average BER Calculation}
By using the series representation of the confluent Hypergeometric function:
\beq
{}_1F_1(a;b;x)=\sum^{\infty}_{n=0}\frac{(a)_n}{(b)_nn!}x^n,
\eeq
where $(\cdot)_n$ is the Pochhammer symbol, in (\ref{final}), we get
\beqarr
f_{\gamma}(\gamma)=A\left(\frac{\eta}{\bar{\gamma}}\right)^{\epsilon}\sum^\infty_{k=0}\tilde{D}_k\sum^\infty_{j=0}C_{j,k}\gamma^{j+\epsilon-1}
e^{-\frac{\eta}{\bar{\gamma}}\gamma},
\label{final1}
\eeqarr
where 
\beq
C_{j,k}=\frac{(Lm+k)_j}{(\epsilon)_jj!}\left(\frac{\eta}{\bar{\gamma}\left(1+\frac{1}{\lambda_1}\right)}\right)^j.
\label{cjk}
\eeq
From the signal-space concept, the average BER of $M$-PSK/QAM constellation is given by~\cite{lu99,bhatn12,arti13a,bhatn14}
\beqarr
Pe(\gamma)=\zeta_M\sum^{\tau_M}_{p=1}Q\left(a_p\sqrt{\gamma}\right),
\eeqarr
where $Q(\cdot)$ is the q-function~\cite[Eq. (2.1.97)]{proak01}, $\zeta_M$, $a_p$, and $\tau_M$ are modulation dependent parameters, given in Table~\ref{tab:1}. The average BER of the considered scheme will be
\beqarr
&&Pe(\bar{\gamma})=A\zeta_M\left(\frac{\eta}{\bar{\gamma}}\right)^{\epsilon}\sum^\infty_{k=0}\tilde{D}_k\sum^\infty_{j=0}C_{j,k}\nn\\
&&\times \sum^{\tau_M}_{p=1}\int^\infty_0\gamma^{j+\epsilon-1}
e^{-\frac{\eta}{\bar{\gamma}}\gamma}Q\left(a_p\sqrt{\gamma}\right)d\gamma.
\label{avgber1}
\eeqarr
\begin{table}
\caption{Values of $\zeta_M$, $\tau_M$, and $a_p$ for different constellations.\vspace*{-3mm}}
\begin{center}
\begin{tabular}{|c|c|c|c|}
\hline
Constellation&$\zeta_M$&$\tau_M$&$a_p$\\
\hline
$M$-QAM&$\frac{4\left(1-\frac{1}{\sqrt{M}}\right)}{\log_2M}$&$\frac{\sqrt{M}}{2}$&$\left(2p-1\right)\sqrt{\frac{3}{(M-1)}}$\\
\hline
$M$-PSK&$\frac{2}{\max\left(\log_2M,2\right)}$&$\max\left(\frac{M}{4},1\right)$&$\sqrt{2}\sin\frac{(2k-1)\pi}{M}$\\
\hline
\end{tabular}
\end{center}\vspace*{-0em}
\label{tab:1}
\end{table}
By using the relation $Q\left(a_p\sqrt{\gamma}\right)=(1/2)\mbox{erfc}(a_p\sqrt{\gamma}/\sqrt{2})$ and \cite[Eq. (2.8.5.7)]{prudn92} in (\ref{avgber1}), we get the average BER of the scheme as
\beqarr
&&Pe(\bar{\gamma})=A\zeta_M\left(\frac{\eta}{\bar{\gamma}}\right)^{\epsilon}\sum^\infty_{k=0}\tilde{D}_k\sum^\infty_{j=0}C_{j,k}\nn\\
&&\times\sum^{\tau_M}_{p=1}\frac{\left(a_p/\sqrt{2}\right)^{-2\epsilon-2j}}{(2\epsilon+2j)\sqrt{\pi}}\Gamma(\epsilon+j+1/2)\nn\\
&&\times{}_2F_1\left(\epsilon+j,\epsilon+j+1/2;\epsilon+j+1;-\frac{2\eta}{a^2_p\bar{\gamma}}\right).
\label{avgber2}
\eeqarr

By employing the relation: ${}_2F_1\left(\epsilon+j,\epsilon+j+1/2;\epsilon+j+1;-\frac{2\eta}{a^2_p\bar{\gamma}}\right)=1$ for very high value of $\bar{\gamma}$ in (\ref{avgber2}), we get
\beqarr
&&Pe(\bar{\gamma})=A\zeta_M\left(\frac{\eta}{\bar{\gamma}}\right)^{\epsilon}\sum^\infty_{k=0}\tilde{D}_k\sum^\infty_{j=0}C_{j,k}\nn\\
&&\times\sum^{\tau_M}_{p=1}\frac{\left(a_p/\sqrt{2}\right)^{-2\epsilon-2j}}{(2\epsilon+2j)\sqrt{\pi}}\Gamma(\epsilon+j+1/2).
\label{avgber3}
\eeqarr

It can be seen from (\ref{cjk}) that $C_{j,k}$ varies inversely with $\bar{\gamma}^j$. Therefore, for the diversity calculation\footnote{The diversity order depends upon the lowest power of the SNR.},
we should set $j=0$. By setting $j=0$ and $p=1$ in (\ref{avgber3}), we get the asymptotic BER of the scheme:
\beqarr
Pe(\bar{\gamma})=\frac{A\zeta_M \Gamma(\epsilon+1/2)}{2\epsilon\sqrt{\pi}\left(a_1/\sqrt{2}\right)^{2\epsilon}}\left(\sum^\infty_{k=0}\tilde{D}_k\right)\left(\frac{\eta}{\bar{\gamma}}\right)^{\epsilon}.
\label{avgber4}
\eeqarr
It is confirmed from (\ref{avgber4}) that the diversity order of the MRC receiver is $\epsilon$, as shown in Subsection~III.C.  
 
Under the condition that $m$ being an integer and $Lm+k>\epsilon$, the Hypergeometric function in (\ref{final}) can be represented
in the form of a finite series by using the Kummer's transform~\cite{abram72} as
\beqarr
&&\hspace*{-2em}{}_1F_1\big(Lm+k;\epsilon;\frac{\eta\gamma}{\bar{\gamma}\left(1+\frac{1}{\lambda_1}\right)}\Big)=e^{\frac{\eta\gamma}{\bar{\gamma}}\left(1+\frac{1}{\lambda_1}\right)^{-1}}\nn\\
&&\hspace*{-2em}\times\sum^{Lm+k-\epsilon}_{j=0}\frac{(Lm+k-\epsilon)!}{(Lm-\epsilon-j)!j!(\epsilon)_j}\left(\frac{\eta\gamma}{\bar{\gamma}\left(1+\frac{1}{\lambda_1}\right)}\right)^j. 
\label{closedform1f1}
\eeqarr
Using (\ref{closedform1f1}), the BER can be calculated in the form of a single power series employing the method stated before. 

\begin{figure}[t!]\vspace*{-1em}
\hspace*{0.0em}
\centerline{\psfig{file=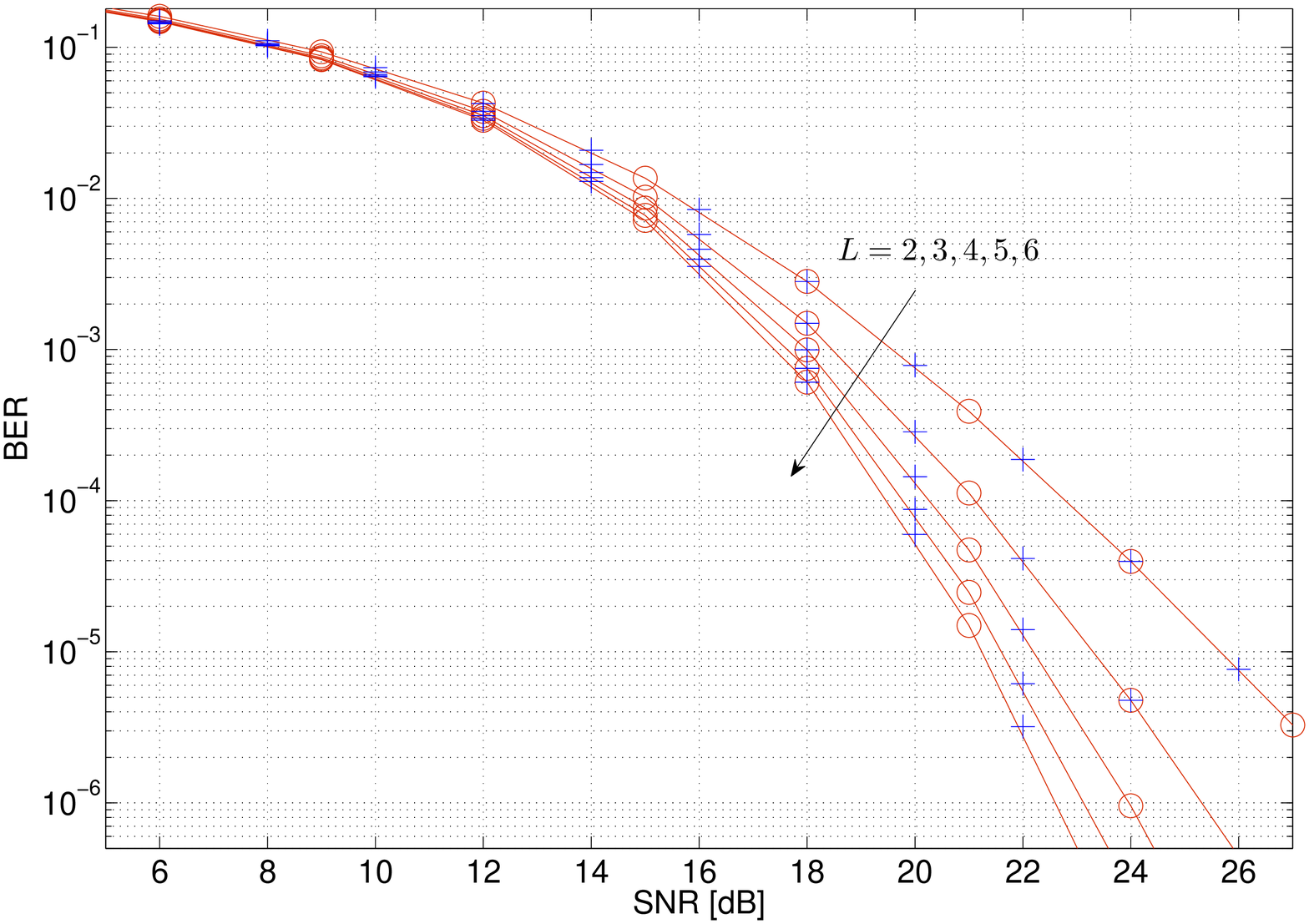,width=4in,height=3.0in}}\vspace*{-4mm}
\caption{Analytical $-\circ-$ and simulated $+$ BERs for $\kappa_l=2$, $\mu_l=2$, $m=2.1$, $\rho_{ij}=0.2^{|i-j|}$, $L=2,3,4,5,6$, and 16-QAM constellation.}\vspace*{-6mm}
\label{fig4}
\end{figure}
The analytical and simulated BER versus SNR plots for $\kappa_l=2$, $\mu_l=2$, $m=2.1$, $\rho_{ij}=0.2^{|i-j|}$, $L=2,3,4,5,6$, and 16-QAM constellation are shown in Fig.~\ref{fig4}. The simulated BER values are obtained by using $10^7$ channel realizations. The closeness of the analytical and simulated BER plots is evident from the figure. Moreover, the performance of the MRC scheme is improved with increasing value of receive antennas, as seen from Fig.~\ref{fig4}.
\subsection{Ergodic Capacity Analysis}
\begin{figure}[t!]\vspace*{-1em}
\hspace*{0.0em}
\centerline{\psfig{file=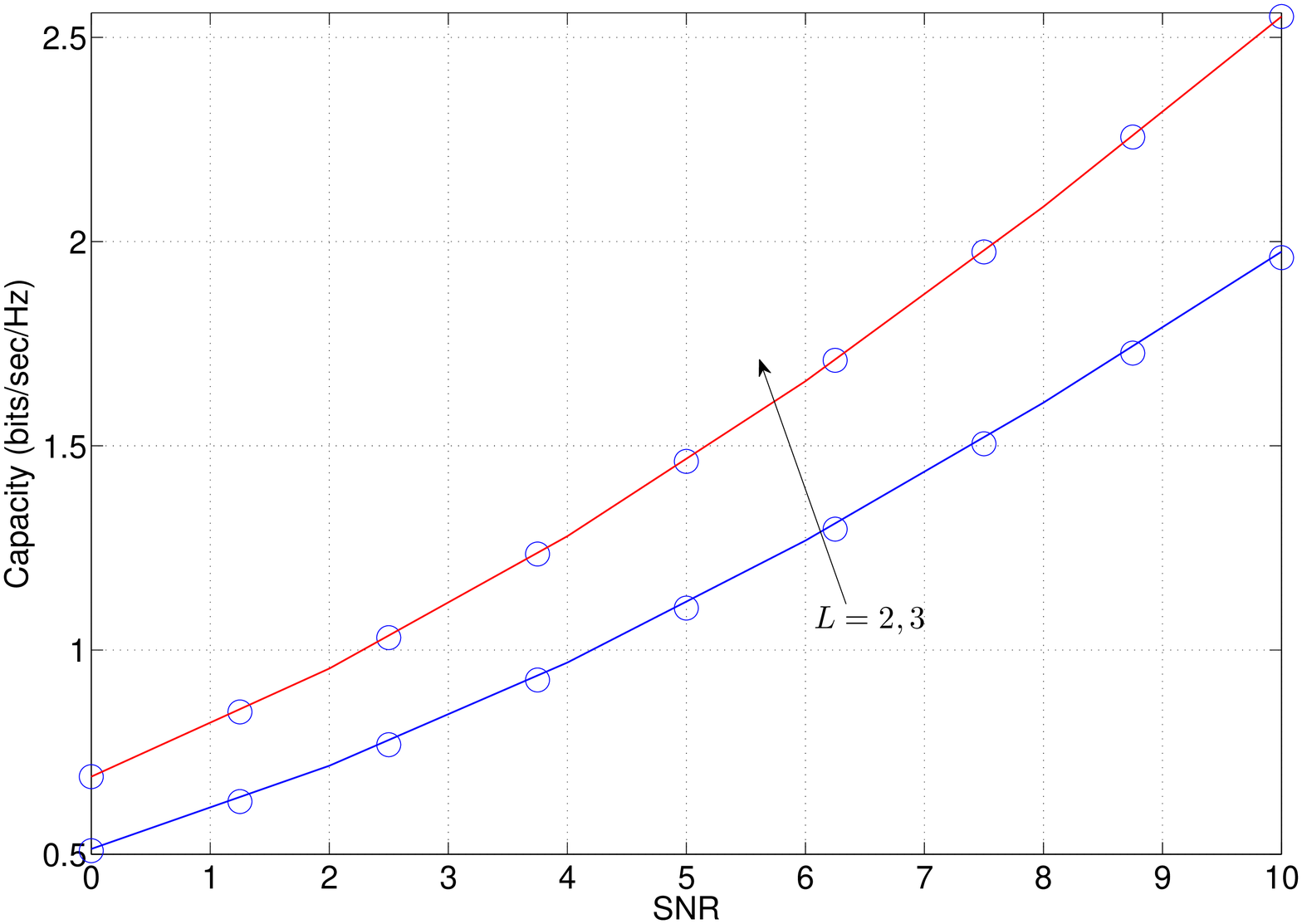,width=4in,height=3.0in}}\vspace*{-4mm}
\caption{Capacity plots for $\kappa_l=2$, $\mu_l=2$, $m=1.2$, $\rho_{ij}=0.5^{|i-j|}$, and $L=2,3$ obtained by using (\ref{cap2}) $-\!\!-$ and (\ref{cap}) $\circ$ .}\vspace*{-6mm}
\label{fig6}
\end{figure}

The average capacity (in bits/sec/Hz) of the scheme is given by
\beq
C(\bar{\gamma})=\int^\infty_0\mbox{log}_2\left(1+\gamma\right)f_\gamma(\gamma)d\gamma.
\label{cap}
\eeq
By using the relations:
   \beqarr
e^{-\frac{\eta}{\bar{\gamma}}\gamma}=G^{10}_{01}\Bigg(\frac{\eta}{\bar{\gamma}}\gamma\Bigg|{.\atop{0}}\Bigg)
\label{expn}
\eeqarr
and
\beqarr
\mbox{ln}\left(1+\gamma\right)=G^{12}_{22}\Bigg(\gamma\Bigg|{\hspace*{-0em}1,1\atop{1,0}}\Bigg),
\eeqarr
where $\mbox{ln}(\cdot)$ is the natural logarithm, and (\ref{exp}) and (\ref{final1}), in (\ref{cap}),
we get
\beqarr
&&C(\bar{\gamma})=\frac{A}{\mbox{ln}2}\left(\frac{\eta}{\bar{\gamma}}\right)^{\epsilon}\sum^\infty_{k=0}\tilde{D}_k\sum^\infty_{j=0}C_{j,k}\nn\\
&&\times \int^\infty_0\gamma^{j+\epsilon-1}
G^{10}_{01}\Bigg(\frac{\eta}{\bar{\gamma}}\gamma\Bigg|{.\atop{0}}\Bigg)G^{12}_{22}\Bigg(\gamma\Bigg|{\hspace*{-0em}1,1\atop{1,0}}\Bigg)d\gamma.
\label{cap1}
\eeqarr
Employing \cite[Eq. (21)]{adamc90} in (\ref{cap1}), we have
\beqarr
&&C(\bar{\gamma})=\frac{A}{\mbox{ln}2}\sum^\infty_{k=0}\tilde{D}_k\sum^\infty_{j=0}C_{j,k}\left(\frac{\eta}{\bar{\gamma}}\right)^{-j}\nn\\
&&\hspace*{3em}\times G^{13}_{32}\Bigg(\frac{\bar{\gamma}}{\eta}\Bigg|{\hspace*{-0em}1,1,1-\epsilon-j\atop{1,0}}\Bigg).
\label{cap2}
\eeqarr
In Fig.~\ref{fig6}, the analytical average capacity of the MRC scheme is plotted for $\kappa_l=2$, $\mu_l=2$, $m=1.2$, $\rho_{ij}=0.5^{|i-j|}$, and $L=2,3$ by using (\ref{cap2}).
Moreover, the plots of the capacity obtained by numerical
integration of (\ref{cap}) are provided in the figure to verify the validity of the proposed capacity
results of (\ref{cap2}). 
It can be seen from the figure that there is a significant improvement in the average capacity of the 
correlated $\kappa-\mu$ shadowed channels with increasing number of the spatial dimension. For example, at SNR=10 dB there is 
30\% increment in the average capacity by using three antennas as compared to the two antennas case. 
%\section{Numerical Results}
%\subsection{Discussion of the Error Rate}
%In Fig.~\ref{}, we have plotted the analytical and simulated values of the SER of the MRC scheme with $\mu_l=\mu=$
\section{Conclusions}
We have statistically characterized the correlated $\kappa-\mu$ shadowed fading in this paper, in terms of PDF and MGF.
These characterizations have been found useful for study of the error performance and diversity performance of
the MRC scheme. The analytical results have been obtained in terms of power series of the Hypergeometric functions and
Meijer-G function. A study of the LOS correlation on the error performance of the MRC receiver
has been performed by using the derived analytical results. It has been deduced on the basis of the analytical results
that the error performance of the receiver is adversely affected by the antenna correlation. 
However, the diversity order of the MRC scheme remains independent of the antenna correlation. 
\bibliography{IEEEabrv,biblitt}

% Generated by IEEEtran.bst, version: 1.13 (2008/09/30)
\begin{thebibliography}{10}
\providecommand{\url}[1]{#1}
\csname url@samestyle\endcsname
\providecommand{\newblock}{\relax}
\providecommand{\bibinfo}[2]{#2}
\providecommand{\BIBentrySTDinterwordspacing}{\spaceskip=0pt\relax}
\providecommand{\BIBentryALTinterwordstretchfactor}{4}
\providecommand{\BIBentryALTinterwordspacing}{\spaceskip=\fontdimen2\font plus
\BIBentryALTinterwordstretchfactor\fontdimen3\font minus
  \fontdimen4\font\relax}
\providecommand{\BIBforeignlanguage}[2]{{%
\expandafter\ifx\csname l@#1\endcsname\relax
\typeout{** WARNING: IEEEtran.bst: No hyphenation pattern has been}%
\typeout{** loaded for the language `#1'. Using the pattern for}%
\typeout{** the default language instead.}%
\else
\language=\csname l@#1\endcsname
\fi
#2}}
\providecommand{\BIBdecl}{\relax}
\BIBdecl

\bibitem{yacou07}
M.~D. Yacoub, ``The $\kappa-\mu$ and the $\eta-\mu$ distribution,'' \emph{IEEE
  Antennas and Propagation Magazine}, vol.~49, pp. 68--81, Feb. 2007.

\bibitem{abdi03}
A.~Abdi, W.~Lau, M.-S. Alouini, and M.~Kaveh, ``A new simple model for land
  mobile satellite channels: {F}irst and second order statistics,'' \emph{IEEE
  Trans.\ Wireless Commun.}, vol.~2, no.~3, pp. 519--528, May 2003.

\bibitem{bhatn13a}
M.~R. Bhatnagar and {Arti M.K.}, ``Performance analysis of {AF} based hybrid
  satellite-terrestrial cooperative network over generalized fading channels,''
  \emph{IEEE Commun.\ Lett.}, vol.~17, no.~10, pp. 1912--1915, Oct. 2013.

\bibitem{mk14a}
{Arti M.K.} and M.~R. Bhatnagar, ``Beamforming and combining in hybrid
  satellite-terrestrial cooperative systems,'' \emph{IEEE Commun.\ Lett.},
  vol.~18, no.~3, pp. 483--486, March 2014.

\bibitem{bhatn13b}
M.~R. Bhatnagar and {Arti M.K.}, ``Performance analysis of hybrid
  satellite-terrestrial {FSO} cooperative system,'' \emph{IEEE Photon. Technol.
  Lett.}, vol.~25, no.~22, pp. 2197--2200, Nov. 2013.

\bibitem{mk14b}
{Arti M.K.} and M.~R. Bhatnagar, ``Two-way mobile satellite relaying: {A}
  beamforming and combining based approach,'' \emph{IEEE Commun.\ Lett.},
  vol.~18, no.~7, pp. 1187--1190, July 2014.

\bibitem{bhatn14}
M.~R. Bhatnagar and {Arti M.K.}, ``On the closed-form performance analysis of
  maximal ratio combining in {Shadowed-Rician} fading {LMS} channels,''
  \emph{IEEE Commun.\ Lett.}, vol.~18, no.~1, pp. 54--57, Jan. 2014.

\bibitem{paris14}
J.~F. Paris, ``Statistical characterization of $\kappa-\mu$ shadowed fading,''
  \emph{IEEE Trans.\ Vehicular Technol.}, vol.~63, no.~2, pp. 518 -- 526, Feb.
  2014.

\bibitem{alfan07}
G.~Alfano and A.~D. Maio, ``Sum of squared {Shadowed-Rice} random variables and
  its application to communication systems performance prediction,'' \emph{IEEE
  Trans.\ Wireless Commun.}, vol.~6, no.~10, pp. 3540--3545, Oct. 2007.

\bibitem{proak01}
J.~G. Proakis, \emph{Digital Communications}, 4th~ed.\hskip 1em plus 0.5em
  minus 0.4em\relax Singapore: McGraw-Hill, 2001.

\bibitem{aloun01}
M.-S. Alouini, A.~Abdi, and M.~Kaveh, ``Sum of {G}amma variates and performance
  of wireless communication systems over {Nakagami}-fading channels,''
  \emph{IEEE Trans.\ Vehicular Technol.}, vol.~50, no.~6, pp. 1471 -- 1480,
  Nov. 2001.

\bibitem{mosch85}
P.~G. Moschopoulos, ``The distribution of the sum of independent {Gamma} random
  variables,'' \emph{Ann. Inst. Statist. Math. (Part A)}, vol.~37, pp.
  541--544, 1985.

\bibitem{dhung12}
Y.~Dhungana, N.~Rajatheva, and C.~Tellambura, ``{Performance analysis of
  antenna correlation on LMS-based dual-hop AF MIMO systems},'' \emph{IEEE
  Trans.\ Vehicular Technol.}, vol.~61, no.~8, pp. 3590--3602, Oct. 2012.

\bibitem{prudn92}
A.~P. Prudnikov, Y.~A. Brychkov, and O.~I. Marichev, \emph{Integrals and
  Series}, 3rd~ed.\hskip 1em plus 0.5em minus 0.4em\relax {New York, USA}:
  Gordon and Breach Science Publishers, 1992, vol.~2.

\bibitem{grand00}
I.~S. Gradshteyn and I.~M. Ryzhik, \emph{Table of Integrals, Series, and
  Products}, 6th~ed.\hskip 1em plus 0.5em minus 0.4em\relax San Diego, CA, USA:
  Academic Press, 2000.

\bibitem{adamc90}
V.~S. Adamchik and O.~Marichev, ``The algorithm for calculating integrals of
  hypergeometric type function and its realization in reduce system,'' in
  \emph{Proc. International Symposium on Symbolic Algebraic Computation
  (ISSAC'90)}, Tokyo, Japan, Aug. 1990, pp. 212--224.

\bibitem{mckay09}
M.~McKay, A.~Zanella, I.~Collings, and M.~Chiani, ``Error probability and
  {SINR} analysis of optimum combining in {Rician} fading,'' \emph{IEEE Trans.\
  Commun.}, vol.~57, no.~3, pp. 676--687, March 2009.

\bibitem{prudn90}
A.~P. Prudnikov, Y.~A. Brychkov, and O.~I. Marichev, \emph{Integrals and
  Series}, 1st~ed.\hskip 1em plus 0.5em minus 0.4em\relax New York, USA: Gordon
  and Breach Science Publishers, 1990, vol.~3.

\bibitem{abram72}
M.~Abramowitz and I.~A. Stegun, \emph{Handbook of Mathematical
  Functions}.\hskip 1em plus 0.5em minus 0.4em\relax New York, USA: Dover
  Publications, Inc., 1972.

\bibitem{lu99}
J.~Lu, K.~B. Letaief, J.~C.-I. Chuang, and M.~L. Liou, ``${M}$-{PSK} and
  ${M}$-{QAM BER} computation using signal-space concepts,'' \emph{IEEE Trans.\
  Commun.}, vol.~47, no.~2, pp. 181--184, Feb. 1999.

\bibitem{bhatn12}
M.~R. Bhatnagar, ``Performance analysis of a path selection scheme in multi-hop
  decode-and-forward protocol,'' \emph{IEEE Commun.\ Lett.}, vol.~16, no.~12,
  pp. 1980--1983, Dec. 2012.

\bibitem{arti13a}
{Arti M.K.} and M.~R. Bhatnagar, ``Performance analysis of hop-by-hop
  beamforming and combining in {DF MIMO} relay system over {N}akagami-$m$
  fading channels,'' \emph{IEEE Commun.\ Lett.}, vol.~17, no.~11, pp.
  2080--2083, Nov. 2013.

\end{thebibliography}
\bibliographystyle{IEEEtran}
%\newpage
\end{document}